\documentclass[aps,floatfix,prd,nofootinbib,superscriptaddress,reprint,showpacs,10pt,preprintnumbers,longbibliography]{revtex4-2}
\usepackage[utf8]{inputenc}
\usepackage[pdftex]{graphicx}
\usepackage{float}
\usepackage{amsmath}
\usepackage{amssymb}
\usepackage{mathtools}
\usepackage{siunitx}
\usepackage{amsfonts}
\usepackage{dsfont}
\usepackage{array}
\usepackage{bm}
\usepackage{mathrsfs}
\usepackage{pifont}
\usepackage{multirow}
\usepackage{upgreek}
\usepackage[dvipsnames]{xcolor}
\usepackage[pdftex,
  pdftitle={Put PDF title here},
  pdfauthor={Put ODF authors here},
  bookmarks,
  colorlinks,
  linkcolor=myblue,
  citecolor=mymagenta,
  menucolor=black,
  urlcolor=myblue,
  plainpages=false,
  pdfpagelabels,
  hypertexnames=false]{hyperref}
\usepackage{verbatim}
\usepackage{slashed}
\usepackage{cleveref}
\usepackage{ucs}
\usepackage{subfigure}
\usepackage{longtable,booktabs}
\usepackage{bbold}

\renewcommand{\l}{\left}
\renewcommand{\r}{\right}
\newcommand{\gev}{\,\mathrm{GeV}}
\newcommand{\mev}{\,\mathrm{MeV}}
\newcommand{\fm}{\,\mathrm{fm}}
\newcommand{\order}[1]{\mathcal{O}\l({#1}\r)}

\newcommand{\hpione}{h^{1}_{\pi}}

\definecolor{mymagenta}{RGB}{200, 0, 100}
\definecolor{myblue}{RGB}{45, 48, 146}

\newcommand{\Llagrange}{\mathcal{L}}

\newcommand{\qbar}{\bar{q}}
\newcommand{\sbar}{\bar{s}}
\newcommand{\tauthree}{\tau^{3}}
\newcommand{\gammafive}{\gamma_{5}}

\newcommand{\brackets}[1]{\langle #1 \rangle}

\renewcommand{\refeq}[1]{(\ref{#1})}
\newcommand{\pvec}{\vec{p}}
\newcommand{\xvec}{\vec{x}}
\newcommand{\yvec}{\vec{y}}
\newcommand{\Tr}[1]{\mathrm{Tr}\left[ #1\right]}
\newcommand{\epow}[1]{\mathrm{e}^{#1}}
\newcommand{\ubar}{\bar{u}}
\newcommand{\dbar}{\bar{d}}
\newcommand{\matelem}{\mathcal{M}}
\newcommand{\cdf}{\mathrm{CDF}}

\newcommand{\MSbar}{\overline{\mathrm{MS}}}
\newcommand{\Saction}{\mathcal{S}}

\newcommand{\psibar}{\bar{\psi}}
\newcommand{\Dop}{\mathcal{D}}
\newcommand{\Pop}{\mathcal{P}}
\newcommand{\Rop}{\mathcal{R}}
\newcommand{\Top}{\mathcal{T}}
\newcommand{\Cop}{\mathcal{C}}
\newcommand{\Sop}{\mathcal{S}}

\graphicspath{{plots/}}

\begin{document}
\title{Exploring a new approach to Hadronic Parity Violation from Lattice QCD}
\author{Marcus Petschlies}
\affiliation{Helmholtz-Institut f\"ur Strahlen- und Kernphysik, University of Bonn, Nussallee 14-16, 53115 Bonn, Germany}
\affiliation{Bethe Center for Theoretical Physics, University of Bonn, Nussallee 12, 53115 Bonn, Germany}
\author{Nikolas Schlage}
\affiliation{Helmholtz-Institut f\"ur Strahlen- und Kernphysik, University of Bonn, Nussallee 14-16, 53115 Bonn, Germany}
\affiliation{Bethe Center for Theoretical Physics, University of Bonn, Nussallee 12, 53115 Bonn, Germany}
\author{Aniket Sen}
\affiliation{Helmholtz-Institut f\"ur Strahlen- und Kernphysik, University of Bonn, Nussallee 14-16, 53115 Bonn, Germany}
\affiliation{Bethe Center for Theoretical Physics, University of Bonn, Nussallee 12, 53115 Bonn, Germany}
\author{Carsten Urbach}
\affiliation{Helmholtz-Institut f\"ur Strahlen- und Kernphysik, University of Bonn, Nussallee 14-16, 53115 Bonn, Germany}
\affiliation{Bethe Center for Theoretical Physics, University of Bonn, Nussallee 12, 53115 Bonn, Germany}

\date{\today}

\begin{abstract}
  The long-range, parity-odd nucleon interaction generated by single pion exchange is captured in the
  parity-odd pion-nucleon coupling $\hpione$. Its calculation in lattice QCD requires the evaluation 
  of 4-quark operator nucleon 3-point functions. We investigate a new
  numerical approach to compute $\hpione$  based on 
  nucleon matrix elements of parity-even 4-quark operators and related to the parity-violating electro-weak
  theory by PCAC and chiral perturbation theory.
  This study is performed
  with 2+1+1 dynamical flavors of twisted mass fermions at pion mass $m_{\pi} \approx 260 \mev$ 
  in a lattice box of $L \approx 3 \fm $ and with a lattice spacing of $a \approx 0.091 \fm$.
  From a calculation excluding fermion loop diagrams we find a bare coupling of
$\hpione =  8.08 \,(98) \cdot 10^{-7}$.

\end{abstract}

\maketitle

\section{Introduction}
\label{seq:introduction}

Determining the effects of hadronic parity violation (HPV) in nucleon-nucleon interaction 
is a challenging task, both in experiment and theory. HPV amplitudes based on parity symmetry breaking
are small deviations against a large QCD background. 
The long-range, single pion exchange interaction
is captured at the hadronic level by the parity-violating Lagrangian
of proton ($p$), neutron ($n$) and the pion triplet ($\vec \pi$)
\begin{align}
  \mathcal{L}^w_{\mathrm{PV}} &= -\frac{\hpione}{\sqrt{2}}\,\bar N\,\left( \vec\tau \times \vec\pi \right)^3
  = i\,\hpione\,\left( n\bar \,p\,\pi^- - \bar p\,n\,\pi^+ \right)\,,
  \label{eq:ia}
\end{align}
which defines the coupling $\hpione$.

It originates from flavor-conserving, neutral currents at the electro-weak scale 
and
is a promising channel to study the parity-odd pion-nucleon coupling~\cite{Guo:2018aiq}.
The first experimental determination~\cite{NPDGamma:2018vhh} of the associated pion-nucleon coupling related to the $\Delta I = 1$
effective electro-weak Lagrangian has recently sparked new interest in
the theoretical Standard Model (SM) prediction of $\hpione$.

The available theoretical estimates of $\hpione$ from SM physics are
predominantly based on effective field theory and model
calculations, apart from one exploratory lattice calculation. 

%

Starting point for model calculations was the scheme for describing parity-nonconserving
nuclear forces from Desplanques, Donoghue, Holstein~\cite{DESPLANQUES1980449}.
Continuing on that basis Dubovik and Zenkin found in Ref.~\cite{Dubovik:1986pj} a best value estimate 
of $1.3\cdot  10^{-7}$. Ref.~\cite{Feldman:1991tj} extended the quark model picture
to include the weak interaction effects from the $\Delta$ baryon and
estimated $2.7\cdot 10^{-7}$.  
Kaiser and Meißner started investigations with a chiral soliton model~\cite{Kaiser:1988bt,Kaiser:1989ah,Kaiser:1989fd},
and Meißner and Weigelt used a three-flavor Skyrme model 
and calculated the coupling in the range  $(0.8 - 1.3)\cdot 10^{-7}$ in Ref.~\cite{Meissner:1998pu}.
A chiral quark-soliton model was used by Ref.~\cite{Hyun:2016ddn}
with an estimate of the coupling $0.874\cdot 10^{-7}$.
Ref.~\cite{Lobov:2002xb} applied the operator product expansion
to the nucleon 2-point function in an external pion field and based on QCD-sum rules
found a value $3.4\cdot 10^{-7}$. Ref.~\cite{Phillips:2014kna} studied
the parity-odd couplings in the nucleon-nucleon interaction with 
the $1/N_c$-expansion and esitmated for the $\sin^2(\theta_w) / N_c$-suppressed
$\hpione$ a range $0.8\,(0.3)\cdot 10^{-7}$. de Vries et al. used chiral
effective field theory in Ref. \cite{deVries:2015pza,deVries:2015gea}
to compute the neutron capture on the proton process and matched to 
experimental data, resulting in an estimate $1.1 \,( 1.0 ) \cdot 10^{-6}$.
The first attempt at an estimate from first principles of 
the strong interaction was carried out by Wasem in Ref.~\cite{Wasem:2011tp}.
We come back to comparing our present work to this reference
and only collect its final estimate here $1.099\,(0.505)\cdot 10^{-7}$.
The significant experimental result by the NDPgamma collaboration
in Ref.~\cite{NPDGamma:2018vhh} of
$2.6\,(1.2)\cdot 10^{-7}$ is another milestone in this timeline.
The same experimental data was subsequently re-analyzed with 
chiral effective field theory in Ref.~\cite{deVries:2020iea},
which estimated the coupling at $2.7\,(1.8)\cdot 10^{-7}$.
More recently, Ref.~\cite{Gardner:2022mxf} used 
a factorization ansatz for the matrix element of
the parity-violating 
electro-weak Hamiltonian, together with non-perturbative
lattice QCD data for the nucleon quark charges
to find an estimate of $3.06\,(1.72)\cdot 10^{-7}$.
 \begin{figure}[htpb]
  \centering
  \includegraphics[width=0.45\textwidth]{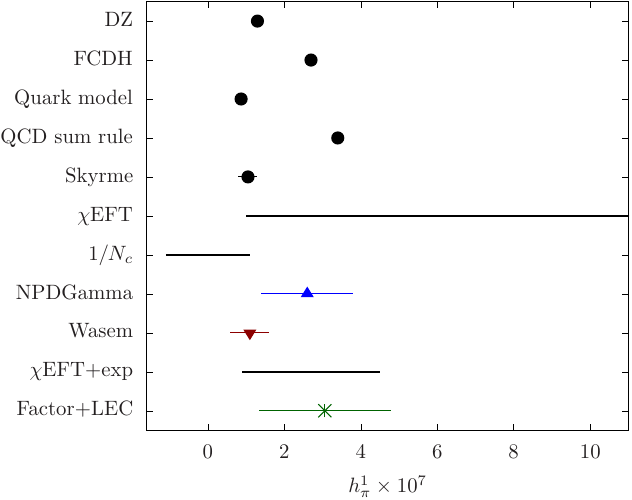}
  \caption{ Collection of estimates of the nucleon-pion coupling $\hpione$; extension of Fig. 4 in ~\cite{NPDGamma:2018vhh}
  and Table I in ~\cite{Guo:2018aiq}.
    The labels match to references: DZ~\cite{Dubovik:1986pj}, FCDH~\cite{Feldman:1991tj}, Quark model~\cite{Hyun:2016ddn}, 
    QCD sum rule~\cite{Lobov:2002xb}, Skyrme~\cite{Meissner:1998pu}, $1/N_c$ ~\cite{Phillips:2014kna}, 
    $\chi$EFT ~\cite{deVries:2015pza}, $\chi$EFT + exp~\cite{deVries:2020iea},
    NPDGamma~\cite{NPDGamma:2018vhh}, 
    LQCD~\cite{Wasem:2011tp}, Factor+LEC~\cite{Gardner:2022mxf}.
    The estimate for $\chi$EFT 2015 extends beyond the shown $\hpione$ axis range $1.1\,(1.0) \cdot 10^{-6}$.
  }
  \label{fig:hpione-comp}
\end{figure}
The above results are summarized in Fig.~\ref{fig:hpione-comp}.
We find this situation of scattered results highlights the need for
systematic study and improved ab-initio theoretical determinations.

The aforementioned first ab-initio lattice QCD determination and the
non-perturbative estimate of the nucleon matrix elements  with the
parity-violating (PV) effective Lagrangian has been presented in
Ref.~\cite{Wasem:2011tp}. In this work the actual transition matrix
elements $N \pi \overset{\Llagrange_{\mathrm{PV}}} {\longrightarrow} N$
for a  transition of a nucleon-pion state to a nucleon state mediated
by the $\Delta I = 1$, parity-violating Lagrangian was
considered. Though this calculation is pioneering, it is
also exploratory in many regards as also discussed in detail in
Refs.~\cite{Wasem:2011zz,Guo:2018aiq}. Challenges are the rigorous
treatment of the pion-nucleon  state in finite volume and energy
non-conservation between initial and finial state on the lattice,
which were circumvented in Ref.~\cite{Wasem:2011tp}. 
Apart from this the calculation considered also only a certain quark
flow diagram topology, arguing that the neglected diagrams are
expected to have a contribution only within the statistical
accuracy. Moreover, renormalization of the 4-quark operators was not
included.  Still, the obtained value on a coarse lattice with a
heavier-than-physical pion $m_{\pi} \approx 390\mev$ is consistent 
with the recent NPDGgamma experimental analysis.

An alternative theoretical ansatz has been put forward anew in
Refs.~\cite{Feng:2017iqb,Guo:2018aiq},
by proposing a joint effort of chiral effective field theory ($\chi$EFT)
and lattice QCD. Based on the PCAC relation the transition via the
parity-violating interaction Lagrangian with a soft pion in the
initial or final state  is equivalent to a transition via a
parity-conserving Lagrangian without a soft pion. 
Details on the relevant PCAC relation
  \begin{align}
    \lim\limits_{p_{\pi}\to 0} \,\brackets{n\,\pi^+\,|\,\mathcal{L}^{w}_{\mathrm{PV}}(0)\,| \,p }
    &\approx 
    -\frac{\sqrt{2}\,i}{F_{\pi}}\,\brackets{p\,|\,\mathcal{L}^w_{\mathrm{PC}}\,|p}
    \label{eq:pcac-1}
  \end{align}
  are provided in Refs.~\cite{Feng:2017iqb,Guo:2018aiq}.
The PCAC relation Eq.~\refeq{eq:pcac-1} is,
however, exact only in the limit of exact chiral symmetry. At
non-zero pion mass it receives higher order corrections in
$\chi$EFT. But these corrections can be argued to be numerically small \cite{Guo:2018aiq},
at the level of $\order{1\%}$ at the physical pion mass,
and can in principle also be calculated by studying the
pion mass dependence with lattice QCD, based on known low-energy constants
from meson and heavy-baryon chiral perturbation theory \cite{Feng:2017iqb}.

This alternative theoretical ansatz leads to a major simplification in
the lattice computation: one now considers a transition amplitude
between single nucleon states $N  \overset{\Llagrange_{\mathrm{PC}}}
{\longrightarrow} N $ with a parity-conserving (PC) Lagrangian, 
which from a numerical point of view is more straightforward to handle
in a lattice calculation. In particular, the complication arising from
the pion-nucleon state is absent since the matrix element is computed
for single nucleon initial and final states.

In this work we investigate the computational concepts proposed in
Ref.~\cite{Guo:2018aiq} in practice and propose a concrete numerical
implementation to evaluate the nucleon 3-point functions with the
4-quark operator insertions of $\Llagrange_{\mathrm{PC}}$. 
Of course the ensuing Wick contractions still comprise fermion loop
diagrams, which were neglected in the first work~\cite{Wasem:2011zz}. We will argue that these diagrams and the
renormalization procedure are intricately linked: in the lattice
calculation these particular fermion loop diagrams generate
power-divergent mixing with lower-dimensional operators, and we add an
initial discussion about such power divergent terms and our future
strategy for renormalizing the 4-quark operators.  \\

Preliminary results of this work have been reported in \cite{Sen:2021dcb,Schlage:2023ljc}.

\section{Operators and coupling}
\label{seq:operators}

The matching between the parity-violating interaction in the
electro-weak sector of the SM and the effective nucleon and pion 
degrees of freedom at energy scale $\Lambda_{\mathrm{QCD}}\sim
m_{\mathrm{proton}}$ has been worked out in Ref.~\cite{Dai:1991bx}.  
Here, we largely follow the notation of the recent
Ref.~\cite{Guo:2018aiq}. The $\Delta I = 1$, parity-conserving
Lagrangian is given by 
\begin{align}
  \Llagrange^{w}_{\mathrm{PC}} &= -\frac{G_F}{\sqrt{2}}\,\frac{\sin^2\left( \theta_w \right)}{3}\,\sum\limits_{i}\,\left( C^{(1)}_i\,\theta^{(\ell)\prime}_i + S^{(1)}_{i}\,\theta^{(s)\prime}_i \right) \,,
  \label{eq:L-PC}
\end{align}
$C^{(1)}$ and $S^{(1)}$ denote the Wilson coefficients obtained in
1-loop perturbation theory~\cite{Dai:1991bx,Kaplan:1992vj}. \\ 
The parity-even 4-quark operators $\theta^{(\ell)\prime}$ with only light quarks
$\ell$ contributing and $\theta^{(s)\prime}$ with light and strange
$s$ quarks contributing read
\begin{align}
  \theta^{(\ell)\prime}_1 &= \qbar_a\,\gamma_{\mu}\,\mathbb{1}\,q_a \, \qbar_b\,\gamma^\mu\,\tauthree \, q_b \,,
  \nonumber \\
  \theta^{(\ell)\prime}_2 &= \qbar_a\,\gamma_{\mu}\,\mathbb{1}\,q_b \, \qbar_b\,\gamma^\mu\,\tauthree \, q_a \,,
  \nonumber \\
  \theta^{(\ell)\prime}_3 &= \qbar_a\,\gamma_{\mu}\,\gammafive\,\mathbb{1}\,q_a \, \qbar_b\,\gamma^\mu\gammafive\,\tauthree \, q_b \,,
  \nonumber \\
  & \nonumber \\
  \theta^{(s)\prime}_1 &= \sbar_a\,\gamma_{\mu}\, s_a \, \qbar_b\,\gamma^\mu\,\tauthree \, q_b \,,
  \nonumber \\
  \theta^{(s)\prime}_2 &= \sbar_a\,\gamma_{\mu}\,s_b  \, \qbar_b\,\gamma^\mu\,\tauthree \, q_a \,,
  \nonumber \\
  \theta^{(s)\prime}_3 &= \sbar_a\,\gamma_{\mu}\,\gammafive\, s_a \, \qbar_b\,\gamma^\mu\gammafive\,\tauthree \, q_b\,,
  \nonumber \\
  \theta^{(s)\prime}_4 &= \sbar_a\,\gamma_{\mu}\,\gammafive\, s_b \, \qbar_b\,\gamma^\mu\gammafive \,\tauthree \, q_a \,.
  \label{eq:4q-op}
\end{align}

In the operator list Eq.~\refeq{eq:4q-op} $q = (u, d)^T$ is the up and down quark doublet and $\tau^3$ is the third Pauli matrix, which gives the iso-vector combination $\bar{u}u - \bar{d}d$.
As a perturbative addition to pure QCD, the interaction Lagrangian
Eq.~\refeq{eq:L-PC} induces a proton-neutron 
mass splitting $(\delta m_N)_{4q}$ due
to the 4-quark operators,
\begin{align}
  \left( \delta m_N \right)_{4q} 
  &= \frac{1}{m_N}\,\brackets{p\,\vert\,\Llagrange^{w}_{\mathrm{PC}}(0)\,\vert\,p}
  = -\frac{1}{m_N}\,\brackets{n\,\vert\,\Llagrange^{w}_{\mathrm{PC}}(0)\,\vert\,n} \,,
  \label{eq:deltamN}
\end{align}
and with the PCAC relation the leading contribution to the coupling comes from the mass splitting
\begin{align}
  \hpione \approx -\frac{1}{F_\pi}\,\frac{(\delta m_N)_{4q}}{\sqrt{2}} \,,
  \label{eq:coupling}
\end{align}
where $F_{\pi}$ denotes the pion decay constant in the chiral limit.

In the following this work is mainly concerned with the lattice QCD estimate of the operator matrix elements 
$\brackets{N\,\vert\,\theta^{(f)}_i(0)\,\vert\,N}$ for $f = \ell,\,s$ and $N$ the proton and neutron.

\section{4-quark operator matrix elements from the lattice}

To determine the nucleon matrix elements of the 4-quark operators in Eq.~\refeq{eq:4q-op} we follow
the Feynman-Hellmann-Theorem
technique advocated for in
Ref.~\cite{Bouchard:2016heu}. The relevant correlation functions
result from inserting the individual operators
$\theta^{(f)\prime}_i(x)$ into the proton or neutron 2-point function,
summed over all lattice sites $x$.

The nucleons are interpolated by the usual zero-momentum, positive parity  proton and neutron 3-quark operators
\begin{align}
  N^{+}_\alpha(t) &= P^{(+)}_{\alpha\alpha'}\, \sum\limits_{\yvec}\,\epsilon_{abc}\,u_a(t,\yvec)^{T}\,C\gammafive\,d_b(t,\yvec)\,\,u_{\alpha' c}(t,\yvec) 
  \nonumber \\
  N^{0}_\alpha(t) &= P^{(+)}_{\alpha\alpha'}\,\sum\limits_{\yvec}\,\epsilon_{abc}\,d_a(t,\yvec)^{T}\,C\gammafive\,u_b(t,\yvec)\,\,d_{\alpha' c}(t,\yvec) 
  \label{eq:nucleon-operator}
\end{align}
with (positive) parity projector $P^{(+)} = \frac{1}{2}\,\left(
\mathbb{1} + \gamma_0 \right)$ and $C$ the charge conjugation matrix.

\begin{figure}[htpb]
  \centering
  \includegraphics[width=0.5\textwidth]{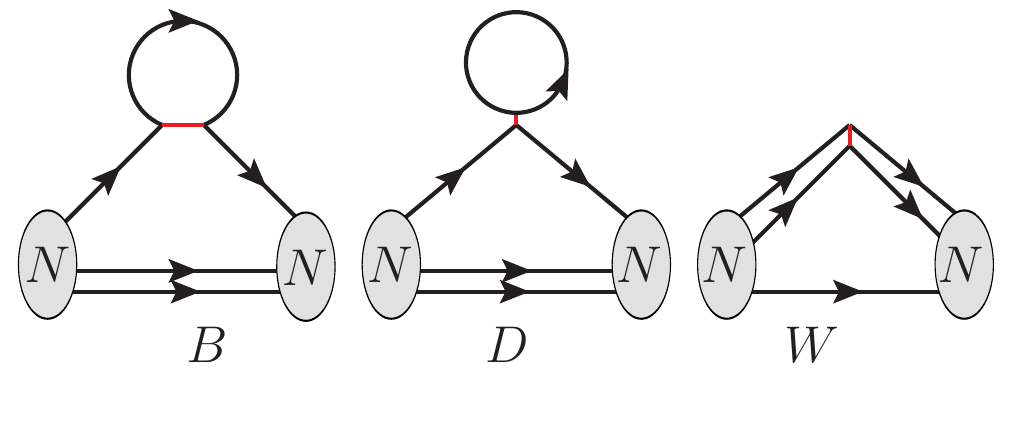}
  \caption{Quark flow diagrams from inserting 4-quark operators into the nucleon 2-point function. 
    The red bar denotes the single-point vertex.
  }
  \label{fig:diagrams}
\end{figure}

From these we construct the 2- and 3-point functions
\begin{align}
  C_{2pt}(t) &= \Tr{ \brackets{ N(t+t_i)\,\,\bar N(t_i) } } \,,
  \label{eq:c2pt} \\
  C_{3pt}(t) &= \sum\limits_{x}\,\, \Tr{ \brackets{ N(t+t_i)\,\, \theta^{(f)\prime}_i(x)\, \,\bar N(t_i) } } \,.
  \label{eq:c3pt}
\end{align}
The different types of Wick contractions following from
Eq.~\refeq{eq:c3pt} are depicted in Fig.~\ref{fig:diagrams} in
diagrammatic form.
We distinguish three types of diagrams: those containing quark loops,
denoted $B$ and $D$, and without a quark loop, denoted $W$. 
Note that the latter type is the only one included in the calculation
of Ref.~\cite{Wasem:2011zz}.  
For the quark loop diagrams, we further make a technical distinction
between type $D$, where the fermion loop is individually spin-color
traced and type $B$, where it is not.

Quark-disconnected diagrams are neglected, since by virtue of the $\tauthree$ flavor structure of the operators in Eq.~\refeq{eq:4q-op} 
such diagrams cancel in SU$(2)$ flavor symmetric QCD, which we work in.

We connect the 2- and 3-point functions in Eqs. \refeq{eq:c2pt}, \refeq{eq:c3pt} to the nucleon matrix element by spectral decomposition and
the Wigner-Eckart-Theorem
\begin{align}
  C_{2pt}(t) &= \sum\limits_{\sigma}\, \epow{-m_N t}  \frac{ \Tr{ \brackets{0\,\vert\,N(0)\,\vert\,n,\sigma} \,\brackets{n,\sigma\,\vert\,\bar N(0)\,\vert\,0} } }{2m_N}
  \nonumber\\
  & \qquad + \dots
  \label{eq:spectral-decomp-1} \\
  C_{3pt}(t) &= (t/a+1)\,\frac{ \brackets{n\,\vert\,\,\theta^{(f)\prime}_i\,\,\vert\,n}}{2m_N} \,\epow{-m_N t} 
  \label{eq:spectral-decomp-2} \\
  & \times \sum\limits_{\sigma}\, \frac{ \Tr{ \brackets{0\,\vert\,N(0)\,\vert\,n,\sigma} \,\brackets{n,\sigma\,\vert\,\bar N(0)\,\vert\,0} } }{2m_N}
  + \dots \,.
  \nonumber
\end{align}
Here $\vert 0 \rangle$ denotes the QCD vacuum state , $\vert n,\sigma \rangle$ the nucleon ground state with zero 3-momentum and 
spin-1/2 component $\sigma$.  In Eq.~\refeq{eq:spectral-decomp-2} we use the spin-independent matrix element 
\begin{align}
   \brackets{n,\,\sigma\,\vert\,\,\theta^{(f)\prime}_i(0)\,\,\vert\,n,\,\sigma'} &= 
   \delta_{\sigma \,\sigma'}\,\brackets{n\,\vert\,\,\theta^{(f)\prime}_i(0)\,\,\vert\,n } \,,
  \label{eq:spectral-decomp-3}
\end{align}
for the Lorentz-scalar operator $\theta^{(f)\prime}_i$.

With ellipsis we denote excited state contributions as well as contributions from different time-orderings, which are
at most of order $\order{t^0}$. A detailed account of the application of the Feynman-Hellmann-Theorem (FHT) 
to the calculation of 
nucleon matrix elements can be found in Ref.~\cite{Bouchard:2016heu}.

According
to FHT, in the vacuum $\vert\lambda\rangle $ including the perturbation $\lambda \,\Llagrange^w_{\mathrm{PC}}$ in the action 
we can determine the effective mass of the nucleon state for sufficiently large $t$ by
\begin{align}
  m_{\mathrm{eff}}^{(\lambda)}(t\,\vert\,\tau) &= 
  \frac{1}{\tau}\,\mathrm{arccosh}\left( \frac{
  C_{2pt}^{(\lambda)}(t+\tau) + 
  C_{2pt}^{(\lambda)}(t-\tau) }{2\,C_{2pt}^{(\lambda)}(t)}
  \right) 
  \nonumber \\
  &= m_{\mathrm{eff}}(t\,\vert\,\tau) + \frac{\lambda}{2}\,(\delta m_N)_{4q} + \order{\lambda^{2}} \,,
  \label{eq:fht-1}
\end{align}
up to excited state contamination, and $m_{\mathrm{eff}}$ as well as the matrix element for $(\delta m_N)_{4q}$ are taken in pure QCD. 

Taking the derivative with respect to $\lambda$ we then obtain the desired matrix element by studying the dependence on source-sink separation
$t$ as well as offset $\tau$ of the ratio
\begin{align}
  R(t,\tau) &=  \frac {\xi}{\sqrt{\xi^2-1} }  \times
  \label{eq:fht-2} \\
  &\quad\times \frac{1}{\tau} \,    \left( 
     \frac{ C_{3pt}(t+\tau) + C_{3pt}(t-\tau) } { C_{2pt}(t+\tau) + C_{2pt}(t-\tau) } - \frac{ C_{3pt}(t) }{C_{2pt}(t) } 
     \right)  \,,
  \nonumber \\
  & \overset{t \mathrm{~large} }{\longrightarrow} \frac{\brackets{n\,\vert\,\theta^{(f)\prime}_i\,\vert\,n} }{2 m_N}
  \nonumber \\
  \xi &= \frac{ C_{2pt}(t+\tau) + C_{2pt}(t-\tau) }{2 C_{2pt}(t) } \,,
  \nonumber
\end{align}
again up to excited state contamination. We determine $R$ per individual operator $\theta^{(f)\prime}_i$ by fitting the ratio
to a constant for various ranges in $t$ and $\tau$.

\begin{center}
  \textit{Evaluation of diagrams}
\end{center}

We evaluate the Wick contractions by a combination of point-to-all, stochastic and sequential quark propagators.
The point-to-all propagators $\psi^{(x_i,\alpha,a)}$ result from solving the lattice Dirac equation for a spin-color diluted source with
support at a single lattice site 
\begin{align}
  S^{\beta\alpha}_{ba}(x; x_i ) &= { D^{-1}}_{\beta,\beta'}^{b,b'}(x,y) \, \left[ \delta_{x_i,y}\,\delta_{\alpha,\beta'}\,\delta_{a,b'} \right] \,.
  \label{eq:pta}
\end{align}
From the point-to-all propagators the nucleon 2-point functions are
evaluated in the usual way.

The quark loop in diagrams $B$ and $D$ in Fig.~\ref{fig:diagrams} is constructed by a fully time, spin and color diluted stochastic timeslice
propagator,
\begin{align}
  L(x)^{ab}_{\alpha\beta} &= \sum\limits_{t,\gamma,c}\, {D_u^{-1} }_{\alpha\kappa}^{ad}(x;y) \,\,
  \left( \eta(t,\yvec)\,\delta_{t_y,t}\,\delta_{\gamma,\kappa}\, \delta_{d,c} \right) 
    \times 
  \nonumber \\
  & \qquad \times \left( \eta(t,\xvec) \, \delta_{t_x,t}\, \delta_{\gamma, \beta}\, \delta_{b,c} \right) \,,
  \label{eq:loop}
\end{align}
where $\eta(t,\xvec) \,\in \, \left\{ \pm 1 \right\}$ are independent and identically with zero mean and unit variance
\begin{align}
  \mathrm{E}\left[ \eta(t,\xvec)  \right] &= 0 \,,\quad 
  \mathrm{E}\left[ \eta(t,\xvec) \,\eta(t,\yvec) \right] = \delta^{(3)}_{\xvec,\yvec} \,.
  \label{eq:binary-source}
\end{align}

To apply the FHT
method we must sum the 3-point function with insertion of the 4-quark operator at each lattice site.
We realize this summed simultaneous insertion by using the sequential inversion method: to that end we construct the two 
sequential sources for $B$- and $D$-type
\begin{align}
  S^{(B)}(x; x_i) &=  \Gamma \,L(x)\,\Gamma\,S(x; x_i) \,,
  \label{eq:B-seq-src} \\
  S^{(D)}(x; x_i) &= \Tr{ \Gamma \,L(x) }\,\Gamma\,S(x; x_i) \,.
  \label{eq:D-seq-src}
\end{align}
Here $\Gamma$ is one of the relevant Dirac matrices $\Gamma = \gamma_\mu$ or $\Gamma = \gamma_\mu\gammafive$. The Lorentz index
$\mu$ is actually summed over at this stage.

By repeated inversion of the Dirac operators on these sources we obtain the sequential propagators
$T^{(B)}$ and $T^{(D)}$ for $B$ and $D$ diagram, respectively, given by
\begin{align} 
  T^{(K)}(y; x_i) &= \sum\limits_{x}\,{ D^{-1} }(y; x) \, S^{(K)}(x; x_i) \,,\quad K = B,\,D \,.
  \label{eq:B-seq}
\end{align}
The ensuing contractions for $C_{3pt}$ are analogous to those for $C_{2pt}$, using $T^{(B)}$ and $T^{(D)}$.

For the $W$ diagram in Fig.~\ref{fig:diagrams} we again use the
stochastic sequential propagator technique to split the four quark lines
connecting at the 
insertion point into two pairs. The relevant term of propagators through the insertion point then reads
\begin{align}
  & \sum\limits_{x}\, S_1(y; x)\,\Gamma\,S_1(x; x_i) \times 
  S_2(y; x)\,\Gamma\,S_2(x; x_i)
  \label{eq:decomp-unity} \\
  &= 
  \sum\limits_{x,z}\, S_1(y; x)\,\Gamma\,S_1(x; x_i) \times 
  S_2(y; z)\,\Gamma\,S_2(z; x_i) \times \delta^{(4)}_{x,z}
  \nonumber \\
  &= \sum\limits_{x,z}\, S_1(y; x)\,\Gamma\,S_1(x; x_i) \times
  S_2(y; z)\,\Gamma\,S_2(z; x_i) \times
  \nonumber \\
  & \qquad \times \mathrm{E}\left[ \eta(x)\,\eta(z)\right] \,,
  \nonumber
\end{align}
with binary noise vector $\eta(x)$ as in Eq.~\refeq{eq:binary-source}, and subscript $1,2$ denoting the quark propagator flavor.

We thus generate a set of independent binary noise sources $\eta^r,\,r=1,\dots,N_r$ as in Eq.~\refeq{eq:binary-source},
and the corresponding sequential sources and propagators
\begin{align}
  S^{(W),r}(x; x_i) &= \Gamma \,\eta^r(x)\, S(x; x_i) \,,
  \nonumber \\
  T^{(W),r}(y; x_i) &= \sum\limits_{x}\,{ D^{-1} }(y; x) \, S^{(W),r}(x; x_i) \,,
  \label{eq:W-seq}
\end{align}
and by Eq.~\refeq{eq:decomp-unity} the product of two such sequential propagators from independent noise
sources produces in the expectation value the four quark lines
connected at a single site, which is summed over the lattice.

\begin{center}
  \textit{Fierz rearrangement}
\end{center}

The operators in Eq.~\refeq{eq:4q-op} fall into two classes with respect to their spin-color structure: 
$\theta^{(f)\prime}_1$ , $\theta^{(f)\prime}_3$  for $f = \ell,\,s$ consist of products of quark bilinear terms, i.e. $\qbar_1 \,\Gamma \, q_1 \times \qbar_2\,\Gamma \,q_2$.
The remaining operators $\theta^{(\ell)\prime}_2$ and $\theta^{(s)\prime}_2,\,\theta^{(s)\prime}_4$ have cross-linked color and spinor indices. We refer to those
as ``color-crossed'' operators for short. \\

These matrix elements of the color-crossed operators can be computed
in two different ways. The first one is to compute the contractions
corresponding to the color-crossed operators. This is achieved 
by using color dilution of the sequential sources $S^{(W),r}$ in Eq.~\refeq{eq:W-seq}:
adding color indices $a,b,c$ we thus have
\begin{align}
  S^{(W),r,a}_{bc}(x; x_i) &= \Gamma \,\eta^r(x)\, S_{ac}(x; x_i)\,\delta_{a,b}\,.
  \label{eq:W-seq-color-diluted}
\end{align}

The second way is to apply a Fierz rearrangement in order to transform
the color-crossed operators into products of standard quark bilinear factors,
thereby avoiding the need for color dilution.
Then, the analogous methods discussed above for the non color-crossed operators
are applied.
For instance $\theta^{(\ell)\prime}_2$ is equivalently represented as
\begin{align}
  \theta^{(\ell)\prime}_2 &= 
  u_{\alpha,a}^{*}\,\left(\gamma_{\mu}\right)_{\alpha\beta}\,u_{\beta,b}\,
  u_{\gamma,b}^{*}\,\left(\gamma_{\mu}\right)_{\gamma\delta}\,u_{\delta,a}  
  - \left[ u \leftrightarrow d \right]
  \nonumber \\
  &= 
  \frac{1}{2}\,\ubar \, \gamma_\mu \, u\,\,\ubar\, \gamma_\mu \, u
  + \frac{1}{2}\,\ubar \, \gamma_\mu \gammafive \, u\,\,\ubar \, \gamma_\mu \gammafive \, u
  \nonumber \\
  &\quad   -\ubar \, \mathbb{1} \, u \,\, \ubar \, \mathbb{1} \, u 
  +\ubar \, \gammafive \, u \,\, \ubar \, \gammafive \, u 
  - \left[ u \leftrightarrow d \right] \,,
  \label{eq:Fierz-1}
\end{align}
where by $\left[ u \leftrightarrow d \right]$ we denote the same term as written explicitly, but up replaced by down quark flavor. The strange operators
are treated accordingly. \\

Thus, instead of the list of operators in Eq.~\refeq{eq:4q-op}, we
consider only the quark bilinear forms
\begin{align}
  \qbar\,\Gamma\,q\,\,\qbar\,\Gamma\,q \,,\quad 
  \sbar\,\Gamma\,q\,\,\qbar\,\Gamma\,s \,,\quad 
  \qbar\,\Gamma\,q\,\,\sbar\,\Gamma\,s \,,
  \quad q = u,\,d
  \label{eq:Fierz-2}
\end{align}
with as before $\Gamma = \left\{ \gamma_\mu \right\},\,\left\{ \gamma_\mu\gammafive \right\} $,
but also in addition 
$\Gamma = \mathbb{1},\,\gammafive $.

\section{Lattice computation}
\label{sec:lattice-computation}

For our numerical study we use a gauge field ensemble from the Extended Twisted Mass Collaboration~\cite{Alexandrou:2018egz} with dynamical 
up, down, charm and strange quark. The ensemble has a lattice volume
of $32^3\times64$, a lattice spacing of $a \approx 0.091\ \mathrm{fm}$ and,
thus, a spatial lattice extend of $L=3.1\ \mathrm{fm}$. The pion mass
value is $m_\pi = 261(1)\ \mathrm{MeV}$, $m_\pi\cdot L=4$ and the
nucleon mass value is $m_N=1028(4)\ \mathrm{MeV}$. The strange quark
mass is tuned to its physical value.

The simulated action features a light mass degenerate quark doublet of
twisted mass fermions at maximal twist guaranteeing $O(a)$
improvement~\cite{Frezzotti:2003ni}, amended by a
Sheikholeslami-Wohlert ``clover'' term included to reduce residual
$a^2$ lattice artifacts. The charm-strange doublet is again of twisted
mass type, including a quark mass splitting term~\cite{Frezzotti:2003xj}.  
The heavy doublet action is not flavor diagonal, which needlessly complicates the calculation of correlation functions involving strange quarks.
For the present study we thus use a mixed-action approach, by the addition of a doublet of Osterwalder-Seiler (OS) strange quarks~\cite{Frezzotti:2004wz}, analogous
to the light quark doublet.

The bare OS strange quark mass value, which is not critically
important yet for this exploratory investigation, has been tuned such
that the  $\Omega$ baryon mass assumes its physical value.

The lattice action determines symmetry properties in our computation
and these are relevant for the discussion of renormalization and
mixing. We thus reprint the detailed formulas for sea and valence
quark action in the App.~\ref{app:action}.

We summarize the statistics produced with the twisted mass gauge field ensemble for the evaluation of the individual diagrams
from the 4-quark operator insertion in Tab. \ref{tab:status}.
\begin{table}
  \centering
  \begin{tabular}{c|r|r|r}
    Diagram type & $N_{\mathrm{conf}}$ & $N_{\mathrm{sample}}$ & $N_{\mathrm{source}}$ \\
    \hline
    $BD$              & 1262 & 768 & 2 \\
    $W$ Fierz-Id      & 1262 & 8 & 2 \\
    $W$ color-crossed & 604 & 8 & 1 \\
\hline
\hline
  \end{tabular}
  \caption{Summary of statistics produced for the evaluation of the individual diagram types from the 4-quark operators
    inserted into the nucleon 2-point function. $N_{\mathrm{conf}}$ gives the number of independent gauge configurations,
    $N_{\mathrm{sample}}$ the number of stochastic samples and $N_\mathrm{source}$ the number of point
    sources employed in the estimates.
  }
  \label{tab:status}
\end{table}


\begin{center}
  \textit{Contributions from $B$ and $D$ diagram type}
\end{center}
For later discussion it will be valuable to consider the contribution
to the nucleon 3-point functions for the individual operators from the
combined $B$ and $D$ diagram type and  
the $W$-type separately, motivated by the fermion loop present in $B$
and $D$-type diagrams, but not in $W$-type diagrams.

\begin{figure*}[htpb]
  \centering
  \includegraphics[width=0.9\textwidth]{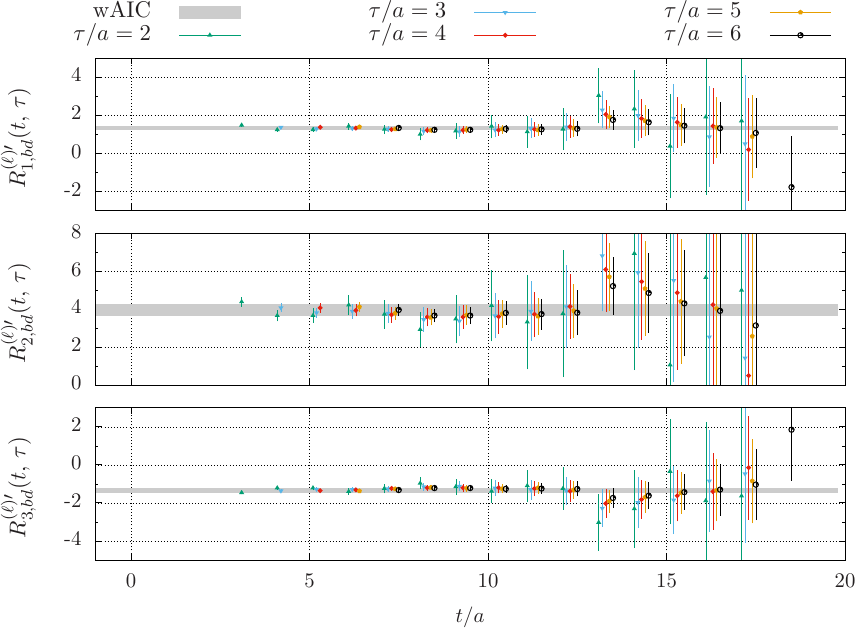}
  \caption{Estimates of $B$ and $D$ diagram contribution to the matrix element from the ratios Eq.~\refeq{eq:fht-2}, 
    for various pairs $(t,\,\tau)$ with the weighted fit as gray
    band. 
  }
  \label{fig:op_bd_rat2_fht_fit}
\end{figure*}

To determine the estimate for the matrix element we fit the ratios $R(t,\tau)$ to a constant in  various ranges 
$t_{\mathrm{min}} \le t \le t_{\mathrm{max}}$ together with various sets $\left\{ \tau_{1},\dots,\tau_{n} \right\}$ of joint data sets. 

The fits are correlated with a block-diagonal covariance matrix, where we neglect the correlation between different $\tau$-values, i.e.
we set $\mathrm{cov}\left( R(t,\tau)\,,\,R(t',\tau') \right) = 0$ for all pairs $\tau \ne \tau'$. Including cross-$\tau$ elements renders the
covariance matrix near-singular and distorts the fit with bad estimates of said elements.

To each fit we assign an Akaike Information Criterion (AIC) weight following the procedure in Ref.~\cite{Borsanyi:2020mff}, which is given by
\begin{align}
  w_{\mathrm{fit}} &= \exp\left(  -\frac{1}{2}\, ( \chi^2  + 2 \,N_{\mathrm{param}} - N_{\mathrm{data}} )  \right)
  \label{eq:aic-1}
\end{align}
with $\chi^2$ based on the block-diagonal covariance matrix, $N_{\mathrm{param}}$ the number of fit parameters (one for our fit to constant),
and $N_{\mathrm{data}}$ the number of data points ( $R(t,\tau)$ values ) entering the fit.

Moreover, the fits are bootstrapped and we obtain the fit parameter uncertainty from the  variance over bootstrap samples.
The ranges for $t$ and $\tau$ applied in the fits are given by
\begin{align}
  1 \le t_{\mathrm{min}} / a \le 10 \,,\quad 
  10 \le t_{\mathrm{max}} / a \le 17 \,,\quad 
  2 \le \tau / a \le 6 \,.
  \label{eq:fit-ranges}
\end{align}
The boundaries are based on observation, where a meaningful fit is accessible, and given our current accuracy the above choice covers all such ranges.
Note in addition, with the symmetric ratio in Eq.~\refeq{eq:fht-2}, the maximal addressed ratio value involves data at $t_{\mathrm{max}}+\tau$.

Finally, we restrict the set of parameters in our fits to a single constant (for the matrix element). At present level of statistical uncertainty per
$R(t,\tau)$ data point, in most $t/\tau$ ranges we cannot model excited state contamination in our data with any statistical significance.

From all available fits with best fit parameter $\mu_i$ and error $\sigma_i$ and AIC weight $w_i$ we build 
the combined distribution function~\cite{Borsanyi:2020mff}
\begin{align}
  \cdf(x) &= \sum\limits_{i}\,w_i\,\mathcal{N}_{\mu_i, \sigma_i}(x) / \sum\limits_{j}\,w_j \,.
  \label{eq:aic-2}
\end{align}
$\mathcal{N}_{\mu,\sigma}$ is the normal distribution with mean $\mu$ and variance $\sigma^2$.
Based on $\cdf(x)$ we quote the median of the distribution function as the central value and the $16 \%$ and $84 \%$ quantiles as the uncertainty interval.

In Fig.~\ref{fig:op_bd_rat2_fht_fit} we present the ratio
$R^{(\ell)'}(t, \tau)$ as a function of $t$ for different values of
$\tau$ for the $B$ and $D$-type operators. In addition we include the
result of the AIC weighting procedure as gray bands. 
The distribution functions and quantiles corresponding to the AIC
procedure are shown in Fig.~\ref{fig:op_bd_rat2_wdf} in App.~\ref{app:aic-wdf}.


\begin{center}
  \textit{Contributions from $W$ diagram type}
\end{center}

The analysis of the $W$ diagram contribution proceeds analogously to the $B+D$ case.

\begin{figure*}[htpb]
  \centering 
  \includegraphics[width=0.9\textwidth]{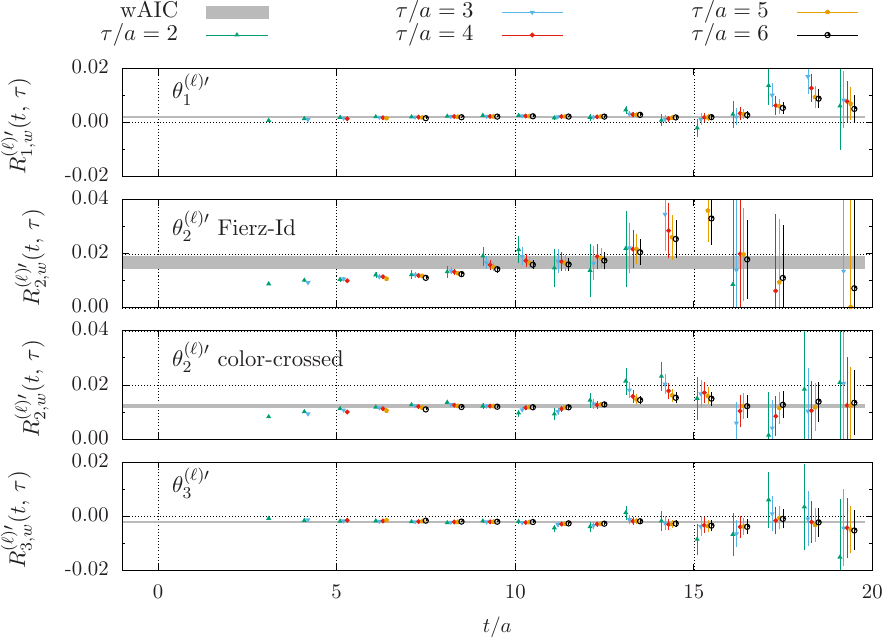}
  \caption{Estimates of $W$ diagram contribution to the matrix element from the ratios Eq.~\refeq{eq:fht-2}; the meaning of symbols as
    Fig.~\ref{fig:op_bd_rat2_fht_fit}.
  }
  \label{fig:op_w_rat2_fht_fit}
\end{figure*}

We show the ratio data $R(t,\tau)$ per operator in Fig.~\ref{fig:op_w_rat2_fht_fit} with our estimate for the matrix element 
as the gray band. Fig.~\ref{fig:op_w_rat2_fht_fit} (App.~\ref{app:aic-wdf}) correspondingly justifies this estimate at the level of the cumulative
distribution function.


\begin{center}
  \textit{Diagrams with strange quarks}
\end{center}

For the strange operators $\theta^{(s)\prime}$ in Eq.~\refeq{eq:4q-op}, only one diagram type contributes per operator.
These are $D$-type contributions for $\theta^{(s)\prime}_{1,\,3}$
and $B$-type contributions for $\theta^{(s)\prime}_{2,\,4}$, the latter
only when using Fierz rearrangement for these two operators. 
Still, whether $B$ or $D$, only strange quark loop diagrams occur in this case.

Here, we circumvent the technical complication of unitary strange and charm flavor mixing by the twisted mass heavy quark action Eq.~\refeq{eq:action_sc}
and employ the aforementioned mixed action approach with a strange quark doublet $(s_{+},\,s_{-})$, analogously to the light quark doublet.

\begin{figure*}[htpb]
  \centering
  \includegraphics[width=0.9\textwidth]{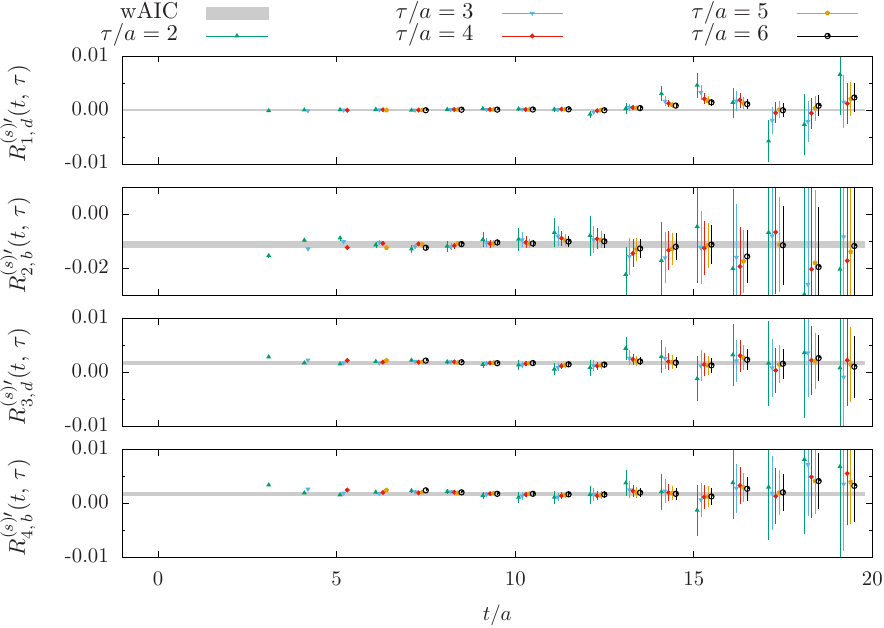}
  \caption{Strange operator ratios $R^{(s)\prime}$ and matrix element fits for operators $\theta^{(s)\prime}$.
  }
  \label{fig:op_strange_rat2_fht_fit}
\end{figure*}

This means the strange operators matrix elements are determined
similarly to the one for $\theta^{(\ell)\prime}$, with
the replacement of the light quark loop by the strange quark loop,
$L_u(x) \to L_s(x)$.
Since the two strange quark flavors in the doublet  $(s_{+},\,s_{-})$
are identical in the continuum limit, we insert
the strange quark loop averaged over both strange quark flavors.
Using $\gammafive$-hermiticity, in detail we then define
\begin{align}
  \bar L_s(x) 
  &= \frac {1}{2}\,\left(  L_{s_+} + L_{s_{-}} \right) 
  = \frac {1}{2}\,\left(  L_{s_+} + \gammafive \,L_{s_+}^\dagger\,\gammafive \right)\,,
  \label{eq:strange-loop} \\
  L_{s_\pm} &= D_{s_{\pm}}^{-1}\,\eta\,\eta^t \,.
  \nonumber
\end{align}
We show the lattice data and the matrix element fit result for the ratios $R^{(s)\prime}(t,\tau)$ 
in Fig.~\ref{fig:op_strange_rat2_fht_fit}. The application of the AIC cumulative distribution functions
defined in Eq.~\refeq{eq:aic-2} is shown in Fig.~\ref{fig:op_strange_rat2_wdf} in App.~\ref{app:aic-wdf}.


\begin{center}
  \textit{Discussion of matrix elements from BDW diagrams }
\end{center}

All results for the matrix elements are compiled in Tabs.~\ref{tab:summary-1} for light quark operators and \ref{tab:summary-2} for the 
strange quark operators. They are given per operator flavor $f = \ell,\,s$, operator number $k$ and as a third label we give the diagrams 
contributing.

\begin{table*}
  \centering
  \begin{tabular}{c||c|c|c||c|c|c}
    \hline
    &&&&&& \\
    $(f,\,k,\,X)$ & 
    $(\ell,\,1, \mathrm{bd})$ &
    $(\ell,\,2, \mathrm{bd})$ &
    $(\ell,\,3, \mathrm{bd})$ &
    $(\ell,\,1, \mathrm{w})$ &
    $(\ell,\,2, \mathrm{w})$ &
    $(\ell,\,3, \mathrm{w})$ \\
    &&&&&& \\
    \hline
    &&&&&& \\
    $\matelem^{(f)\prime}_{k,\,X} / (2a m_N )$ & 
  $  1.307\,_{-0.094}^{+0.082} $ &
  $  3.93 \,_{-0.28}^{+0.25} $ &
  $ -1.320 \,_{-0.085}^{+0.097} $ &
  $ 0.218\,_{-0.017}^{+0.019} \cdot 10^{-2} $ &
$ 1.66\,_{-0.21}^{+0.24}\cdot 10^{-2} $ &
$ -0.218\,_{-0.025}^{+0.022}\cdot 10^{-2} $ \\
&&&&&& \\
\hline
\hline
  \end{tabular}
  \caption{Summary of best estimates for dimensionless bare operator matrix elements per diagram type $B+D$ (``bd'') and $W$ (``w''),
  divided by $2 m_N$, from our fits and AIC analysis.}
  \label{tab:summary-1}
\end{table*}

\begin{table*}
  \centering
  \begin{tabular}{c||c|c|c|c}
    \hline
    &&&& \\
    $(f,\,k,\,X)$ &
    $(s,\,1, \mathrm{d})$ &
    $(s,\,2, \mathrm{b})$ &
    $(s,\,3, \mathrm{d})$ &
    $(s,\,4, \mathrm{b})$ \\
    &&&& \\
    \hline
    &&&& \\
    $\matelem^{(f)\prime}_{k,\,X} / (2a m_N )$ &
$ 0.002\,_{-0.008}^{+0.010} \cdot 10^{-2} $ &
$ -1.117\,_{-0.100}^{+0.107} \cdot 10^{-2} $ &
$ 0.182\,_{-0.024}^{+0.021} \cdot 10^{-2} $ &
$ 0.185\,_{-0.032}^{+0.029} \cdot 10^{-2} $ \\
&&&& \\
\hline
\hline
  \end{tabular}
  \caption{Summary of best estimates for dimensionless bare strange operator matrix elements,
  divided by $2 m_N$, from our fits and AIC analysis.}
  \label{tab:summary-2}
\end{table*}

In Tab.~\ref{tab:summary-1} we observe a difference in magnitude between $B+D$ and $W$ diagram contributions for each individual operator by two orders of
magnitude. Our explanation here mixing with operators of lower (and equal) mass-dimension, in the case of $B$ and $D$ diagrams. The latter two types
contain a quark loop and we argue in Sec.~\ref{sec:mixing} below, that mixing of the 4-quark operator is permitted, starting with local quark-bilinear operators.
By naive visual inspection, the structure of the $W$ diagram on the other hand, does not allow for such mixing and in Sec. \ref{sec:bare-coupling} below
we use solely its contribution to arrive at an estimate of $\hpione$ in analogy to Ref.~\cite{Wasem:2011zz}.

The strange quark operators $\theta^{(s)\prime}_k$ are entirely built from $B$- and $D$-type diagrams,
albeit with the strange quark flavor running inside the fermion loop.
These operators are therefore equally prone to mixing as the light quark operators. 

This mixing of $\left\{ \theta^{(\ell)\prime}_k , \,\theta^{(s)\prime}_k \right\}$ in lattice QCD with lower dimensional operators
does not come entirely unexpected, given the dimension 6 of the operators and the reduced symmetry of the lattice model.
We add several comments on potential subtractions and renormalization in Sec.~\ref{sec:mixing} below.

A second feature is the antisymmetry of $\matelem^{(\ell)\prime}_1$ and $\matelem^{(\ell)\prime}_3$. The opposite-equal values
can be shown for the bare matrix element at tree-level of perturbation theory. Beyond that, we are currently unaware of a symmetry 
argument, that would enforce this property at the non-perturbative level.

\section{Bare coupling from $W$-type diagram}
\label{sec:bare-coupling}

We put our present study in perspective to the work of Ref.~\cite{Wasem:2011tp} by taking an analogous approach: 
we only include the contribution from the $W$-type diagram and ignore
the multiplicative renormalization and the mixing to match the lattice result to 
the $\overline{MS}$ scheme, while still using the Wilson coefficients from renormalized perturbation theory.

Thus, we evaluate the combination of matrix elements $\matelem^{(\ell)\prime}_i = \brackets{N\,\vert\,\theta^{(\ell)\prime}_i\,\vert\,N}/(2 m_{N})$
\begin{align}
  \matelem_{C} &= \sum\limits_{i=1}^{3}\,C^{(1)}_{i}\,\matelem^{(\ell)\prime}_i \,.
  \label{eq:W-1}
\end{align}
Based on the sets of bootstrapped fits with their associated AIC weights, we construct the cumulative distribution function for $\matelem_{C}$
by first building all combinations of fits for $\matelem^{(\ell)}_i$, $i=1,2,3$, then determining the mean and error from bootstrap mean and variance
of the sample-wise built linear combination in Eq.~\refeq{eq:W-1}, and finally assigning to each such combination of fits the AIC weight as the product
of weights from the three individual matrix element fits.
\begin{figure}[htpb]
  \centering
  \includegraphics[width=0.5\textwidth]{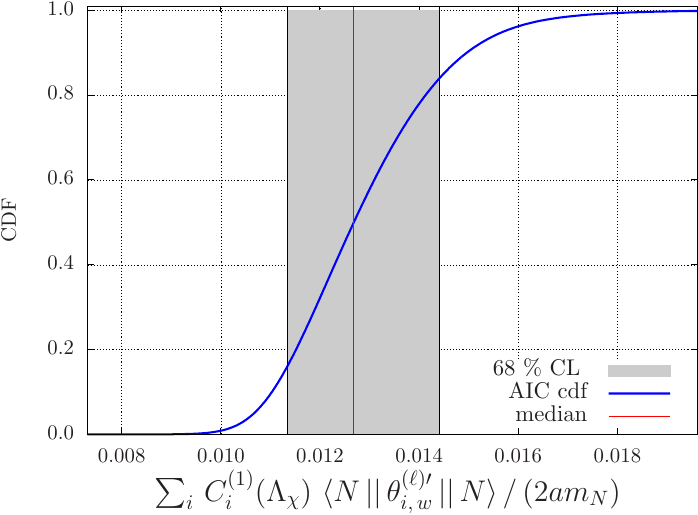}
  \caption{Cumulative distribution function for the combined estimate of $\matelem_C$ in Eq.~\refeq{eq:W-1}.}
  \label{fig:W-cdf}
\end{figure}
The resulting cumulative distribution function is shown in Fig.~\ref{fig:W-cdf}, together with the median ($\alpha = 0.5$ quantile)
and the confidence band from the $\alpha = 0.16 ,\, 0.84$ quantiles.

From the matrix element estimate $\matelem_{C}$ we determine the bare coupling based on the $W$-type diagrams by multiplying the numerical factors
from the effective Lagrangian. 
\begin{align}
  \hpione\left( W,\,\mathrm{bare} \right) &= \frac{G_F\cdot (\hbar c /a)^2\,\sin^2(\theta_W)}{3\,a f_\pi} 
  \nonumber \\
  & \qquad \times \ \frac{a^4\,C^{(1)}_{i}\,\brackets{N\,\vert\,\theta^{(\ell)\prime}_i(0)\,\vert\,N }}{2\,am_N}
  \label{eq:h-1}
\end{align}
In Eq.~\refeq{eq:h-1} we restored all explicit factors of the lattice spacing $a$ to have dimensionless quantities only.\footnote{The nucleon state has mass dimension $-1$, due to normalization $\brackets{N,\pvec\,\vert\,N,\pvec'} = 2E_N(\pvec)\,L^3\,\delta_{\pvec\pvec'}$ in finite volume $L^3$.}

The pion decay constant for the gauge field ensemble considered has been determined 
in Ref.~\cite{ExtendedTwistedMass:2021gbo} with value in lattice units $af_\pi = 0.06674 \,(15)$.
The only explicit use of the lattice spacing is made to convert the
Fermi constant $G_F$ as an external scale to lattice units. We use the value
$a = 0.09076\,(54)\,\mathrm{fm}$ from Ref.~\cite{Alexandrou:2022amy}.

For the matrix elements combined with the Wilson coefficients in lattice units we find
\begin{align}
  \frac{a^4\,C^{(1)}_{i}\,\brackets{N\,\vert\,\theta^{(\ell)\prime}_i(0)\,\vert\,N }}{2\,am_N}
  &= 1.27 \,_{13}^{17} \cdot 10^{-2} \,,
  \label{eq:ci-m}
\end{align}
and together with the conversion factor with Standard Model parameters from PDG~\cite{Workman:2022ynf}
converted to lattice units
\begin{align}
  \frac{G_F\cdot (\hbar c /a)^2\,\sin^2(\theta_W)}{3\,af_\pi} &= 6.367 \, ( 77 ) \cdot  10^{-5}\,,
\label{eq:conversion_factors}
\end{align}
the result for the bare coupling is then
\begin{align}
  \hpione\left( W,\,\mathrm{bare} \right) &= 8.08\,( 98 ) \cdot 10^{-7}\,.
  \label{eq:hpione-estimate}
\end{align}
The result in Eq.~\refeq{eq:hpione-estimate} as representative of a lattice estimate of $\hpione$ is by construction very preliminary.
Conceptually, in its underlying restrictions it is similar to the first lattice determination in Ref.~\cite{Wasem:2011zz}, which found
\begin{align}
  \hpione\left( \mathrm{Wasem~2012} \right) &= (1.099 \pm 0.505)\,^{+0.058}_{-0.064}\cdot 10^{-7}\,,
  \label{eq:hpione-Wasem}
\end{align}
at $m_{\pi} \approx 389\mev$ pion mass and with a coarser lattice $a = 0.123 \fm$. We recall, that the computational ansatz 
of both lattice calculations differs fundamentally in using a parity-violating versus parity-conserving interaction Lagrangian.

Moreover, our preliminary result for $\hpione(W,\,\mathrm{bare})$ is of compatible by order of magnitude
with the recent experimental value
\begin{align}
\hpione(\mathrm{exp}) &= 
2.6\,(1.2)_{\mathrm{stat}}\,(0.2)_{\mathrm{syst}} \cdot  10^{-7}\,.
  \label{eq:hpione-experiment}
\end{align}

\section{Comments on mixing and outline of renormalization}
\label{sec:mixing}

The renormalization of the set of 4-quark operators in Eq.~\refeq{eq:4q-op} in continuum QCD
has been discussed in Refs.~\cite{Dai:1991bx,Kaplan:1992vj}, and finds
its expression in the Wilson coefficients $\left\{ C^{(1)}_i,\,S^{(1)}_i \right\}$  which are calculated in renormalized QCD perturbation
theory in the $\MSbar$ scheme, together with their anomalous dimension matrix. In continuum QCD with the mass-independent $\MSbar$ scheme
and dimensional regularization, mixing with operators of lower dimension is excluded.

Here we comment on the situation in the practical lattice QCD calculation, with Wilson-type fermions and non-zero quark mass. 
The explicit breaking of proper Lorentz symmetry down to discrete 3-rotations and to non-equivalence of spatial and temporal direction (due to $T \ne L$)
does not play a significant role for the 4-quark operators, which are invariant under a 3- and 4-dimensional rotation.

Of practical importance in this numerical calculation with Wilson-type fermions is the breaking of chiral symmetry, of up-down SU(2) flavor
and parity symmetry. We focus in these introductory remarks on the light quark propagators $\theta^{(\ell)\prime}$.

\begin{center}
  \textit{ Twisted Mass fermions}
\end{center}
The symmetries of the 
twisted mass lattice action for the light quarks are listed in Eqs.~\refeq{eq:tra-C}, \refeq{eq:tra-P}, \refeq{eq:tra-T}, \refeq{eq:tra-R5}, \refeq{eq:tra-Dd} and \refeq{eq:tra-Sud} in the App.~\ref{app:action}.

Based on these lattice symmetries we identify the operators which are allowed to mix, i.e. which are not excluded by quantum numbers
under those symmetry transformations. These quantum numbers are for the (light) 4-quark operators
\begin{center}
  \begin{tabular}{l|ccccc}
  Operator                  & $\Pop\Dop[-m_f]$ & $\Top\Dop[-m_f]$ & $\Cop$ & $\Dop\Rop_5$ & $\Pop\Sop$ \\
  \hline
  $\theta^{(\ell) \prime} $ & $+1$             & $+1$             & $+1$   & $+1$         & $-1$ \\
  \hline
\end{tabular}
\end{center}
The mixing candidate operators of mass-dimension 3 to 5 are given by

\begin{center}
  \begin{tabular}{l|l}
  dim   &     \\
  \hline
  3 & $\qbar\,\gammafive \otimes \mathbb{1}\,q$ \\
  \hline
  4 & $m_\ell\,\qbar\,\mathbb{1} \otimes \tauthree \,q$, \,\, $ \qbar\,\slashed{D} \otimes \tauthree \,q$  \\
  \hline
  5 & $m_\ell^2\,\qbar\,\gammafive \otimes \mathbb{1}\,q$ ,\,\,  $m_{\ell}\, \qbar\,\gammafive\slashed{D} \otimes \mathbb{1}\,q$, \\
  & $ \qbar\,\gammafive\,D^2 \otimes \mathbb{1} \,q$, \,\,
  $\qbar \, \sigma_{\mu\nu}\,\tilde G_{\mu\nu}\,q$,\,\, $m_{\ell}\,G\tilde G$ \\
  \hline
\end{tabular}
\end{center}
Here, $\mathbb{1}$ denotes the unit matrix in spinor and flavor space, respectively, and $\tilde G_{\mu\nu} = \epsilon_{\mu\nu\alpha\beta}\,G_{\alpha\beta}$
the dual (lattice) gluon field strength tensor.\footnote{On the same footing there are mixing candidate operators of dimension 6.
  These, however, cause at most logarithmic divergent scaling violations, which we defer to later discussion.}
 The dimension-3 operator is allowed by breaking of parity $\Pop$ symmetry by the twisted mass fermion action. The dimension-4 operators
are allowed by chiral symmetry breaking, which in addition to having opposite $\Rop_5$ parity to $\theta^{(\ell)\prime}$ is indicated
by the explicit factor of the fermion mass $m_{\ell}$. Both operators are connected by the equation of motion for the quark field.
Analogous statements hold for the dimension-5 operators.\\

Mixing with such operator matrix elements hampers the extraction of the renormalized 4-quark operator matrix elements in the continuum limit, due to
power-divergent mixing coefficients $1/a^3$ and $1/a^2$ for dimension-3 and -4 operators, respectively. A suitable scheme to subtract
such contributions appears to be the gradient flow together with the short flow time expansion~\cite{Kim:2021qae}. The practical application
to the twisted mass case is currently under investigation.

We mention two other setups, which are of interest in this study of the method using parity-even 4-quark operators.

\begin{center}
  \textit{ Iso-symmetric Wilson fermions}
\end{center}
Wilson fermions observing SU(2) isospin symmetry have individual parity $\Pop$ and exchange symmetry $\Sop$ (cf. Eqs.~\refeq{eq:tra-C}-\refeq{eq:tra-Sud}).
What remains is the explicit chiral symmetry breaking by the Wilson term. In this case the dimension-3 operator is ruled out by parity symmetry. 
However, the dimension-4 operator $m_{\ell}\,\qbar\,\mathbb{1}\otimes \tauthree \,q$ (and thus the related $\qbar\,\slashed{D}\otimes \tauthree \,q$) 
is still allowed to mix.

\begin{center}
  \textit{Parity-odd 4-quark operators}
\end{center}
This case was investigated in Ref.~\cite{Wasem:2011tp}, with iso-symmetric clover-improved Wilson fermions. In this case, using
parity $\Pop$, charge conjugation $\Cop$ and exchange symmetry $\Sop$ is sufficient to show, that there is no operator of dimension
3, 4 or 5, that can mix on the lattice with the operators of $\Llagrange_{\mathrm{PV}}^w$. \\

The different mixing properties in the lattice QCD calculation with Wilson-type fermion regularization appear as a major drawback of 
using the PCAC relation and converting to the parity-conserving Lagrangian. However, in the parity-violating case
the accurate representation of the nucleon-pion state with
the L\"uscher method and the ensuing signal-to-noise problem from the meson-baryon interpolator pose potentially even harder 
problems, especially towards physical pion mass and large lattice volume.

\section{Conclusion}
\label{sec:conclusion}

In this work we investigated the numerical implementation of a new method to calculate nucleon matrix elements of parity-even
flavor-conserving 4-quark operators from lattice QCD, that pertain to the fully theoretical prediction of the long-range nucleon-pion coupling $\hpione$.
This constitutes the first step towards a full-fledged calculation of the coupling from a combination of chiral perturbation theory 
and non-perturbative lattice matrix elements.
For one ensemble at pion mass $260\mev$ and lattice spacing $a \approx 0.091 \fm$ we demonstrated 
for the first time the calculation of all relevant (bare) matrix elements at the level of $10 \% $ combined statistical and systematic
uncertainty. The specific implementation shown here is readily and feasibly scalable towards physical pion mass, continuum and infinite volume.

We achieve this result due to the simplified representation of the coupling by application of the soft-pion-theorem and the implied sufficiency 
to calculate (single-hadron) nucleon matrix elements of parity-even operators. Of course, corrections to this leading order
defining relation in $\chi$PT, though expected to be small, can in
principle be computed and, therefore, the approximation is
systematically improvable.

However, renormalization of lattice matrix elements poses a challenge due to mixing with lower-dimensional operators. 
\begin{table}
  \centering
  \begin{tabular}{c|c|c}
     $W$ light & $B+D$ light & $B+D$ strange \\
     & & \\
     \hline
     & & \\
     $1.27 \,_{13}^{17} \cdot 10^{-2}$ &
     $4.12 \,_{20}^{20}$ &
     $3.10\,_{29}^{25} \cdot 10^{-2}$ \\
     & & \\
     \hline
     \hline
  \end{tabular}
  \caption{Summary of matrix elements per diagram type, weighted by the Wilson coefficients.}
  \label{tab:matelem}
\end{table}
To appreciate the expected impact of this mixing we summarize our results for the bare matrix elements
normalized by twice the nucleon mass in lattice units and
weighted by the Wilson coefficients in Tab.~\ref{tab:matelem}.
In the table we keep the three contributions
\begin{align}
  & \sum\limits_{i=1}^{3}\,C^{(1)}_i\,\matelem^{(\ell)'}_{i,w} / (2am_{N})\,,\,\,\,\,
  \sum\limits_{i=1}^{3}\,C^{(1)}_i\,\matelem^{(\ell)'}_{i,bd} / (2am_N) \,,
  \nonumber \\
  & \qquad \sum\limits_{i=1}^{4}\,S^{(1)}_i\,\matelem^{(s)'}_{i,bd} / (2am_N)\,,
  \label{eq:cm}
\end{align}
separate, which are light operators with $W$-type diagrams, light operators with $B+D$-type diagrams 
and the light-strange operators with $B+D$-type diagrams. 
From our numerical investigation we find that only $W$ light has the expected order of magnitude, while
the light and strange $B+D$-type diagram contributions are much larger. The latter two types contain 
fermionic loop sub-diagrams, which we deem as the origin of the mixing.

This mixing 
comes about due to reduced symmetries at non-zero lattice spacing and
explicit breaking of chiral symmetry due to finite quark mass values. The Gradient Flow method
for subtracting power-divergent mixing and renormalization appears as
a promising direction to study, due to the possibility to 
study mixing and matching to a continuum renormalization scheme only after extrapolating lattice data to the continuum, and thus
with restored symmetries. The investigation of its practical implementation for the pertinent 4-quark operators is our on-going work.

\begin{acknowledgments}
  We are grateful to Andrea Shindler, Tom Luu and Jangho Kim for useful discussions
  on the subject of renormalization. \\
  This work is supported by the Deutsche
  Forschungsgemeinschaft (DFG, German Research Foundation) and the  
  NSFC through the funds provided to the Sino-German
  Collaborative Research Center CRC 110 “Symmetries
  and the Emergence of Structure in QCD” (DFG Project-ID 196253076 -
  TRR 110, NSFC Grant No.~12070131001)
  The open source software packages tmLQCD~\cite{Jansen:2009xp,Abdel-Rehim:2013wba,Deuzeman:2013xaa}, 
  Lemon~\cite{Deuzeman:2011wz}, 
  QUDA~\cite{Clark:2009wm,Babich:2011np,Clark:2016rdz}, R~\cite{R:2019} 
  and CVC~\cite{cvc} have been used.
\end{acknowledgments}

\appendix
\section{Lattice action and symmetries}
\label{app:action}

\begin{center}
  \textit{Light quark lattice action}
\end{center}

\onecolumngrid
The lattice action of up and down quark for twisted mass fermions with a clover term is given by
\begin{align}
  \Saction^{(\ell)} &= a^4\;\sum\limits_{f=u,d} \; \sum\limits_{x} \;
  \psibar_f(x)\; \left(
    \gamma_\mu\,\bar{\nabla}_\mu  
    - i \gammafive\,  W_{\mathrm{cr}} + m_f 
    + r_f\,\frac{ac_{\mathrm{SW}}}{4}\,\gammafive\,\sigma\,G
  \right) \; \psi(x) \,,
  \label{eq:action_l} 
\end{align}
with Wilson parameters $r_u = 1 = -r_d $.
\twocolumngrid

In Eq.~\refeq{eq:action_l} we have covariant derivative
\begin{align*}
  \bar{\nabla}_{\mu} &= \frac{1}{2}\,\left( \nabla^f_\mu + \nabla^b_\mu \right) \,,
\end{align*}
the subtracted Wilson term of dimension 5
\begin{align*}
  W_{\mathrm{cr}} &= - \frac{a\, r_f}{2} \,\nabla^f_\mu \,\nabla^b_\mu + M_{\mathrm{cr}}(r_f) \,,
\end{align*}
and the Sheikholeslami-Wohlert term  again of dimension 5 and with the clover-plaquette-based lattice field strength tensor $G$~\cite{Sheikholeslami:1985ij}.

The form of the twisted mass lattice action~\refeq{eq:action_l} is valid in the physical basis of the quark fields, i.e. where the 
mass term is real and diagonal, at maximal twist ~\cite{Shindler:2007vp}, such that automatic $\order{a}$ improvement is
realized for physical observables.

\begin{center}
  \textit{Discretization of the strange quark fermion action}
\end{center}

For the strange quark we employ the mixed action technique, with different fermion discretization used for the sea quarks pertaining to
gauge field sampling and to the valence quark, used for the actual calculations of correlation functions.   \\

\onecolumngrid
The sea quark action is given in Ref.~\cite{Frezzotti:2004wz} and to realize a mass splitting and automatic $\order{a}$ improvement 
features mixing strange and charm sea quark flavor by lattice artifacts.
\begin{align}
  \Saction^{(s,c)}_{\mathrm{sea}} &= a^4\; \sum\limits_{x} \;
  \psibar_h(x)\; \left(
    \gamma_\mu\,\bar{\nabla}_\mu  
    - i \gammafive\,\tau^1\,W_{\mathrm{cr}} 
    + \mu_{\sigma} + \mu_\delta\,\tau^3
    + \frac{ac_{\mathrm{SW}}}{4}\,\tau^1\, \gammafive\,\sigma\,G
  \right) \; \psi_h(x) \,,
  \label{eq:action_sc} 
\end{align}
where $\psi_h = (c,\,s)^T$ denotes the strange-charm doublet, $\mu_{\sigma}$ the average bare quark mass of the doublet and $\mu_{\delta}$ the mass
splitting.
The bare mass parameters $\mu_{\sigma},\,\mu_{\delta}$ are tuned by the two conditions of physical $D_s$ meson mass, as well as the
renormalized quark mass ratio $\left. m_s / m_c\right|_{\overline{\mathrm{MS}},\mu = 2\gev}$. \\

\twocolumngrid

To simplify the calculation of nucleon correlators with strange operator insertion we follow the mixed action approach in
Ref.~\cite{Frezzotti:2004wz} and introduce another
doublet of twisted mass valence strange quarks $(s_+,s_-)$. It is formally identical to the light quark doublet, except for
the value of the bare twisted quark mass $m_{\ell} \to m_{s}$.

In particular it shares the critical hopping parameter (as tuned to maximal twist), 
and the Sheikholeslami-Wohlert parameter $c_{\mathrm{SW}}$
with the light quark sector. The bare quark mass $m_s$ at maximal twist is given by twisted quark mass parameter $m_s = \mu_s$ and the latter
is tuned, such that the mass of the $\Omega$ baryon takes the physical value.

\begin{center}
  \textit{Symmetry transformations for light quarks}
\end{center}

We list the complete set of discrete transformations,
which pertain to our identification of operator mixing for the 4-quark operators.
Apart from those listed here, there are the 3-rotations, and the residual (continuous) $\text{U}(1)_{3}$ flavor symmetry, under which the lattice
action is invariant. \\

The discrete transformations of charge conjugation $C$, parity $P$ and time reversal $T$ are given by
\begin{align}
  \psi(t,\xvec)    &\stackrel{C}{\to} C^{-1} \,\psibar(t,\xvec)^T
    \,,\quad
    \psibar(t,\xvec) \stackrel{C}{\to} -\psi(t,\xvec)^T\,C 
    \nonumber \\
    U_\mu(t,\xvec) &\stackrel{C}{\to} U_\mu(t,\xvec)^*
    \label{eq:tra-C} \\
  & \nonumber \\
    \psi(t,\xvec)    &\stackrel{P}{\to} \gamma_4\,\psi(t,-\xvec) 
    \,,\quad
    \psibar(t,\xvec) \stackrel{P}{\to} \psibar(t,-\xvec)\,\gamma_4 
    \nonumber \\
    U_4(t,\xvec)     &\stackrel{P}{\to} U_4(t,-\xvec) 
    \,,\quad
    U_k(t,\xvec)     \stackrel{P}{\to} U_k(t,-\xvec-a \hat k)^\dagger 
  \label{eq:tra-P} \\
  & \nonumber \\
    \psi(t,\xvec)    &\stackrel{T}{\to} i\gamma_4\,\gammafive\,\psi(-t,\xvec) 
    \,,\quad
    \psibar(t,\xvec) \stackrel{T}{\to} -i\psibar(-t,\xvec)\,\gammafive\,\gamma_4 
    \nonumber \\
    U_4(t,\xvec)     &\stackrel{T}{\to} U_4(-t-a,\xvec)^{\dagger}
    \,,\quad
    U_k(t,\xvec)     \stackrel{T}{\to} U_k(-t,\xvec)
  \label{eq:tra-T}
\end{align}
In addition to the discrete Lorentz transformation there are several spurious transformations
\begin{align}
 \mathcal R_5 &: \left\{
 \begin{matrix}
   \psi \to \gammafive\,\psi \,, \quad \psibar \to -\psibar\,\gammafive
 \end{matrix} \right\}
  \label{eq:tra-R5} \\
  \mathcal D_d &: \left\{ 
    \begin{matrix}
      \psi \to \epow{ \frac{3}{2}\pi\,i}\,\psi \,,\quad 
      \psibar \to \epow{ \frac{3}{2}\pi\,i}\,\psibar \\
      U_\mu(x) \to U_{\mu}(-x-a\hat \mu)^{\dagger} \\
    \end{matrix}
  \right\}
  \label{eq:tra-Dd} \\
  \mathcal S_{u,d} &: \left\{ 
    \begin{matrix} u / d \to d / u \,,\quad 
  \ubar / \dbar \to \dbar / \ubar
\end{matrix}
  \right\}
  \label{eq:tra-Sud}
\end{align}
The transformations 
\refeq{eq:tra-C},
\refeq{eq:tra-P},
\refeq{eq:tra-T}, \refeq{eq:tra-R5}, \refeq{eq:tra-Dd} and \refeq{eq:tra-Sud} form a complete set
to define the light quark action and form operator multiplets eligible for mixing. The following transformations are symmetries
\begin{align}
 & P \times \mathcal D_d \times (m_f \to -m_f) \,,
  \label{eq:sym_PDm} \\
  & T \times \mathcal D_d \times (m_f \to -m_f)  \,,
  \label{eq:sym_TDm} \\
  & \mathcal C \,,
  \label{eq:sym_C} \\
  & \mathcal D_d \times \mathcal R_5 \,,
  \label{eq:sym_DR5} \\
  & P \times \mathcal{S}_{u,d} \,,
  \label{eq:sym_PS}
\end{align} 
where the spurious transformation $m_f \to -m_f$ denotes the change of sign of the bare mass parameters.

\section{AIC cumulative distribution functions}
\label{app:aic-wdf}

\begin{figure}[htpb]
  \centering
  \includegraphics[width=0.5\textwidth]{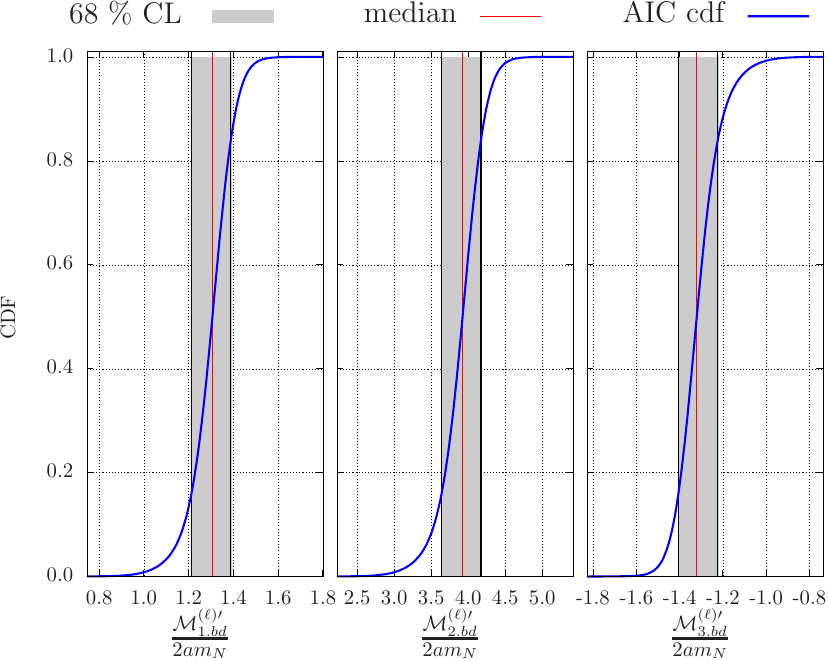}
  \caption{Cumulative distribution function built from AIC weights, for $\theta^{(\ell)\prime}_k$ matrix element $\matelem^{(\ell)\prime}_{k}$;
  the AIC-based cumulative distribution function is shown as the blue line, the ensuing median in red; the gray shaded band is bounded
  by the $16\%$ and $84\%$ quantiles.}
  \label{fig:op_bd_rat2_wdf}
\end{figure}

\begin{figure}[htpb]
  \centering
  \includegraphics[width=0.5\textwidth]{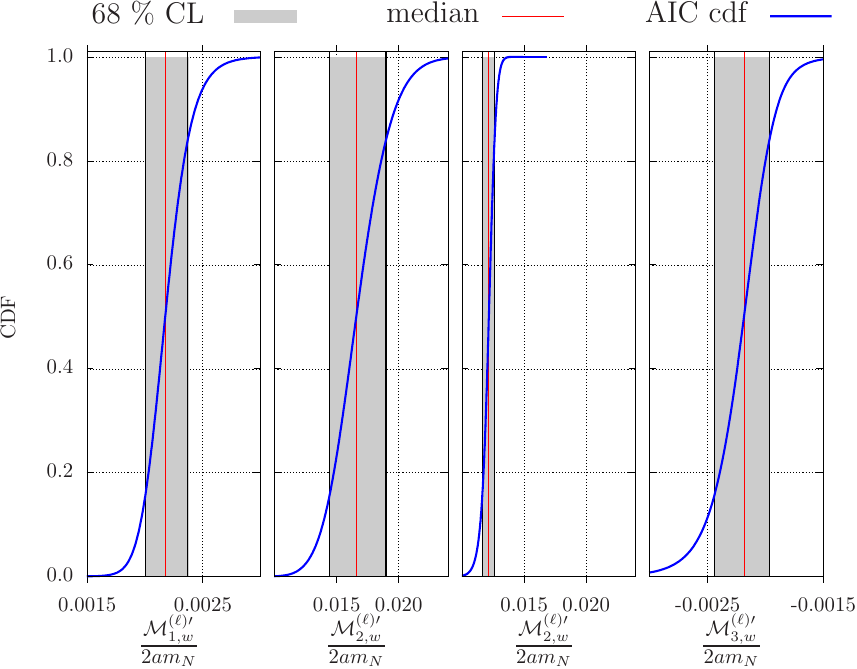}
  \caption{Cumulative distribution function built from AIC weights for $W$ diagram contribution.}
  \label{fig:op_w_rat2_wdf}
\end{figure}

\begin{figure}[htpb]
  \centering
  \includegraphics[width=0.5\textwidth]{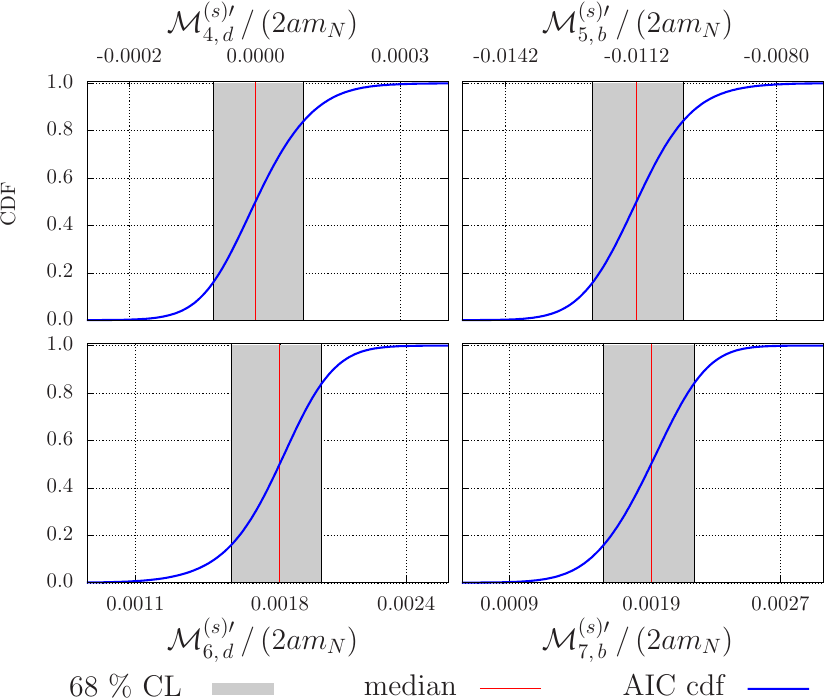}
  \caption{Strange operator AIC analysis by cumulative distribution function.}
  \label{fig:op_strange_rat2_wdf}
\end{figure}

%

\end{document}